\documentclass[shortnote]{jpsj2}
\usepackage {graphicx}
\title{Power Law of Customers' Expenditures in Convenience Stores}

\author{Takayuki Mizuno\thanks{E-mail address: mizuno@ier.hit-u.ac.jp}}

\inst{Institute of Economic Research, Hitotsubashi University, 2-1 Naka, 
Kunitachi City, Tokyo 186-8603}

\kword{Point-of-sale, Expenditure distribution, Power law, 
Convenience store, Econophysics}

\begin{document}
\maketitle

\newpage 

 We can observe a power law in critical phenomena in natural sciences. Such 
a power law is also empirically exhibited in many economic phenomena. In one 
of the older studies conducted approximately 120 years ago, Pareto first 
reported that the probability density function of personal wealth in Italy 
follows a power law [1]. Following this research, the power law was also 
found to hold for city size [2, 3], the circulation of magazines and 
newspapers [4], book sales [5], and the number of incoming links to a 
webpage from another webpage [6, 7]. For the last ten years, investigations 
on this subject particularly focused on the fact that the probability 
density functions of market price fluctuations [8, 9] and company size both 
follow power laws [10--17], based on the analysis of high-frequency data.

 Recently, point-of-sale (POS) databases containing detailed records of all 
customer purchases in shops were published and attracted the attention of 
both physicists and economists. For example, an accelerating power law 
growth for book sales was found in the time series of the book sales on 
Amazon.com [18]. Further, it was reported that approximately 20{\%} of a 
store's products contribute to 80{\%} of its sales [19, 20]. 

 In this paper, we focus on a POS database for convenience stores. First, we 
provide a description of the POS database. Next, we investigate that the 
probability density function of the expenditure of a person during a single 
shopping trip. The top 10 {\%} of the function follows a power law with an exponent of --2.5. 
Moreover, the exponent is independent of the 
store's location, the shopper's age, the day of week, and the time of day.

In this study, we analyze a POS database containing detailed records of all 
purchases at a convenience store chain. The database comprises 
``NEEDS-SCAN/CVC receipt data'' and is published by Nikkei Media Marketing, 
Inc. The convenience store chain sells many foods manufactured by 
Yamazaki Baking Co., Ltd. Therefore, we believe that the name of this chain 
is Daily Yamazaki Co., Ltd.; however, the Nikkei has not officially 
announced the chain's name. The database recorded the receipt number, the 
store location, the purchase date and time, the customer's age and sex, and 
the name of the item for each item purchased. The general location of a 
store is chosen from among the following four categories: in the vicinity of 
a railroad station, a business district, a residential area, and the 
roadside. The customer's age is estimated by the cashier. In the database, 
the customers are categorized into the following four age groups: children, 
youth, adults, and the elderly.

In Fig.1, we present a cumulative density function of the expenditure of a 
person during a single shopping trip by analyzing the 9,750 receipts of the 
people for purchases from June to August 2006. The vertical axis, $P( \ge 
E)$, represents the cumulative probability in log scale, that is, the 
probability of finding a person who spends more than $E$ yen during a single 
shopping trip. We focus on the amount scale that is greater than 900 yen. 
$P( \ge 900yen) \cong 0.1$ is found. As illustrated in Fig.1, the top 10{\%} 
of the expenditure function can be approximated as follows:

\begin{align}
P( \ge E) \propto E^\beta \ \ (\beta = - 2.5),
\end{align}

\noindent
where $E$ is the expenditure of a person during a single shopping trip and 
$\beta $ expresses the power exponent. In Fig.1, we can also find that the 
expenditure function follows the lognormal distribution for $E < 900$ yen. 
Therefore, the function on the bottom 90 {\%} of 
expenditures is approximated by the lognormal distribution. Next, we will 
demonstrate that the exponent of --2.5 is independent of the shopper's age, 
the store's location, the day of week, and the time of day. 

 First, we categorize the receipts by the customers' age. Fig.2(a) presents 
a cumulative density function of the expenditures according to each age 
group. On average, a child spends less than the others in convenience 
stores. For example, the average expenditures of a child and an adult are 
438 yen and 566 yen, respectively. However, the expenditure function follows 
a power law with an exponent of --2.5 that is independent of age. We were 
unable to judge whether the function of elderly people follows the power law 
because their sample size is small.

Next, we focus on the store locations. Many people visit stores located in 
business districts at lunchtime. Customers typically buy two food items and 
one beverage. Meanwhile, with regard to roadside stores, people often travel 
there by car and buy many daily necessities and miscellaneous goods at the 
same time. An average customer at a roadside store spends a considerable 
amount of money in comparison with one who visits a store in a business 
district. Surprisingly, in each store location, the density function of the 
expenditure of a customer during a single shopping trip follows a power law 
with an exponent of -2.5, as illustrated in Fig.2(b). 

In Fig.2(c), we present the expenditure functions from Monday to Tuesday, 
Wednesday to Friday, and holidays. The functions follow a power law with an 
exponent of --2.5 that is independent of the day of week. We believe that 
the exponent does not change from day to day.

Most convenience stores have longer shopping hours, with some being open 24 
hours a day. The average expenditure of a person during a single shopping 
trip depends on the time periods of day. The average expenditures are 400 
yen in the morning, 550 yen during the daytime, and 750 yen at midnight. In 
Fig.2(d), we present the cumulative density function of the expenditure of a 
person during a single shopping trip for these three time periods. We can 
approximate a tail of the function by a power law with an exponent of --2.5 
that is independent of the time periods.

 It is found that a tail of the 
cumulative density function of the expenditure of a person during a single 
shopping trip follows a power law with an exponent of --2.5 that is 
independent of the store's location, the shopper's age, the day of the week, 
and the time of day. Such a power law was also empirically exhibited in many 
economic phenomena. We are interested in the universal dynamics of the 
economic phenomena that follow the power law. A POS database containing 
detailed records of purchases may be able to facilitate our understanding of 
the universal dynamics.

\newpage 
\section*{Acknowledgment}

I would like to thank Misako Takayasu for the many discussions held at 
various stages of this work, and Eri Amemiya for carefully reading this 
paper. This work is partly supported by Research Fellowships of the Japan 
Society for the Promotion of Science for Young Scientists.

\newpage 

\section*{Figure Captions}

\noindent
\\[12pt]
{\bf Fig.1} Distribution of the expenditure of a person during a single shopping 
trip (diamond plots). For $E \ge 900$ yen, the 
distribution follows a power law with an exponent of --2.5. For $E < 900$ yen, 
the distribution is approximated by a 
lognormal distribution with $<$\textit{logE}$>$ = 2.60 and \textit{$\sigma $}(\textit{logE}) = 0.33, where 
\textit{logE} indicates the logarithmic expenditure.

\noindent
\\[12pt]
{\bf Fig.2} Expenditure distribution in each case. The guidelines follow a power 
law with an exponent of --2.5. (a) The expenditure distributions of children, youth, adults, and the 
elderly. (b) The expenditure distributions in roadside areas, residential areas, 
business districts, and in the vicinity of railroad stations. (c) The expenditure distributions from Monday to Tuesday, Wednesday to 
Friday, and holidays. (d) The expenditure distributions from 08:00 to 09:59, 10:00 to 17:59, and 
19:00 to 06:59.

\newpage 

\begin{figure}
\centerline{\includegraphics[height=3in]{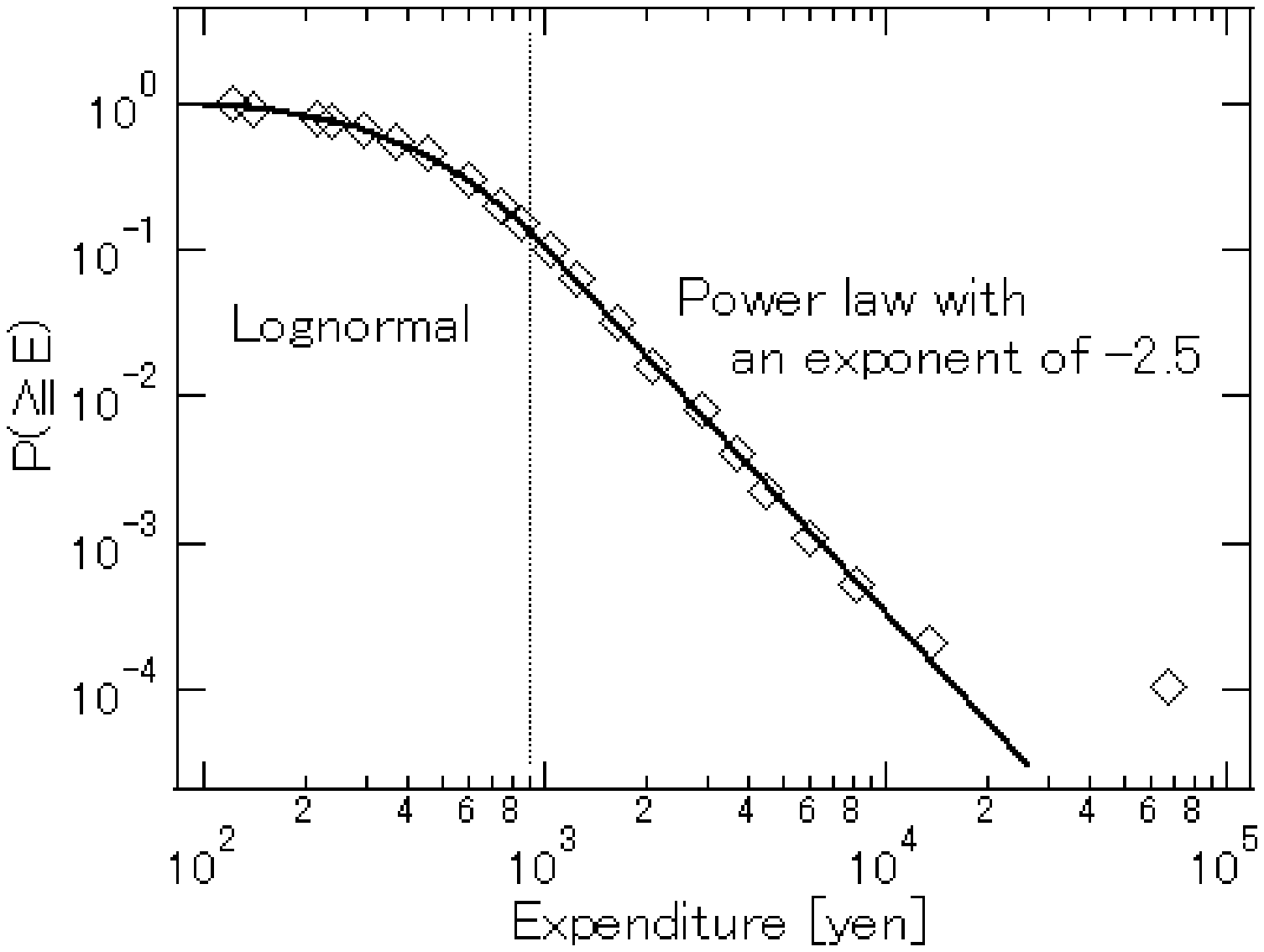}}
\caption{}
\end{figure}

\newpage 
\begin{figure}
\centerline{\includegraphics[height=2.2in]{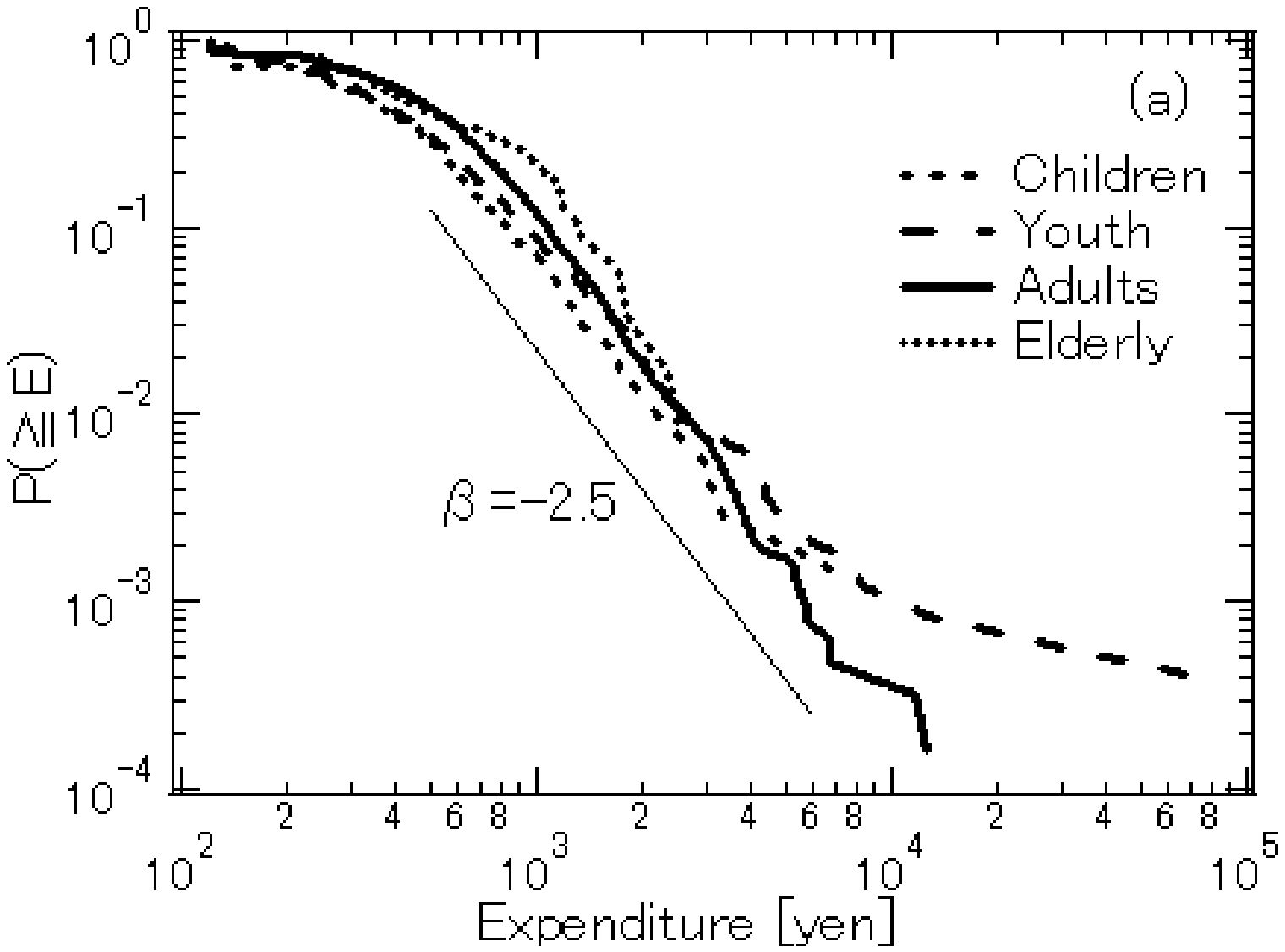}}
\centerline{\includegraphics[height=2.2in]{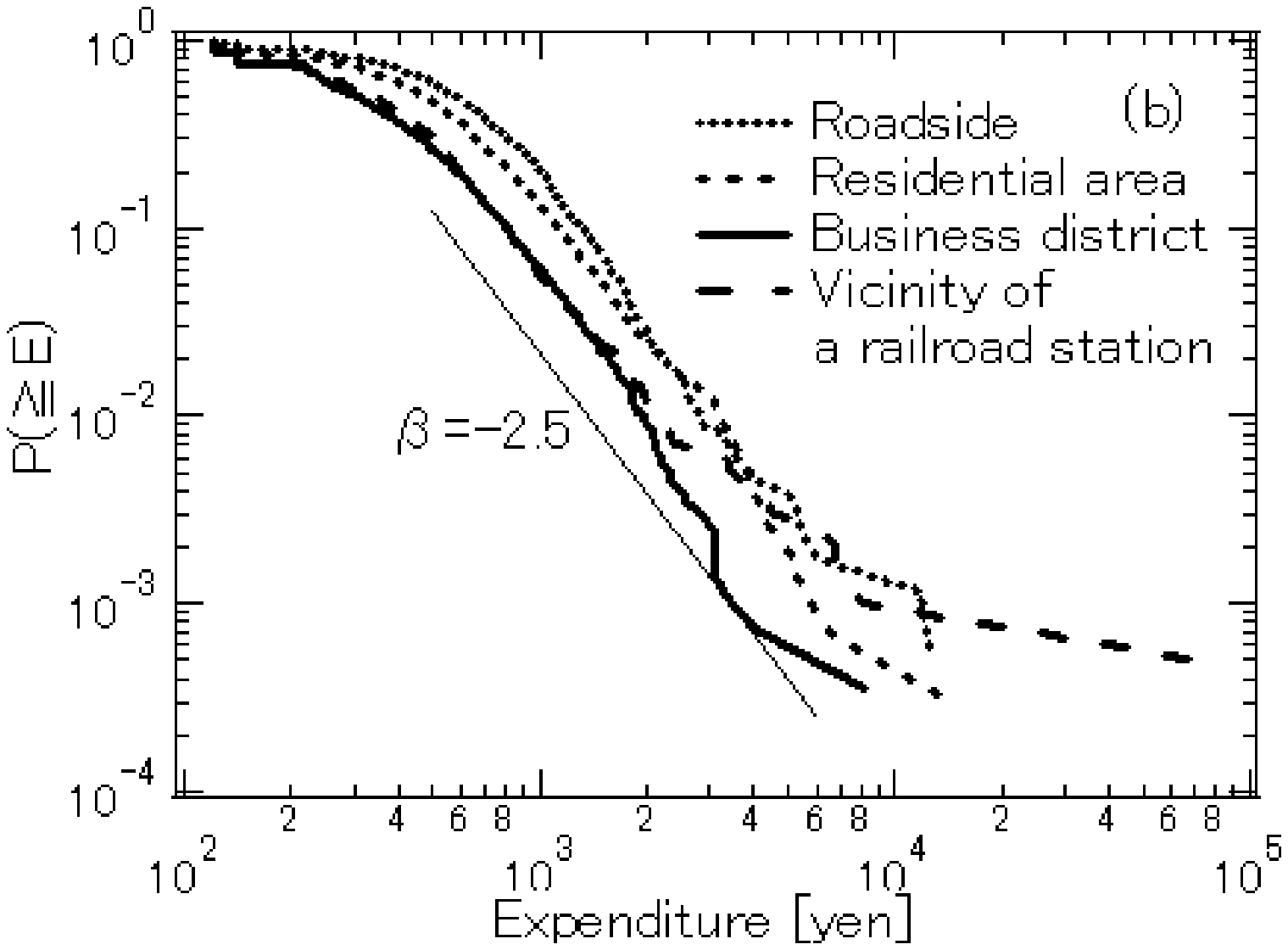}}
\centerline{\includegraphics[height=2.2in]{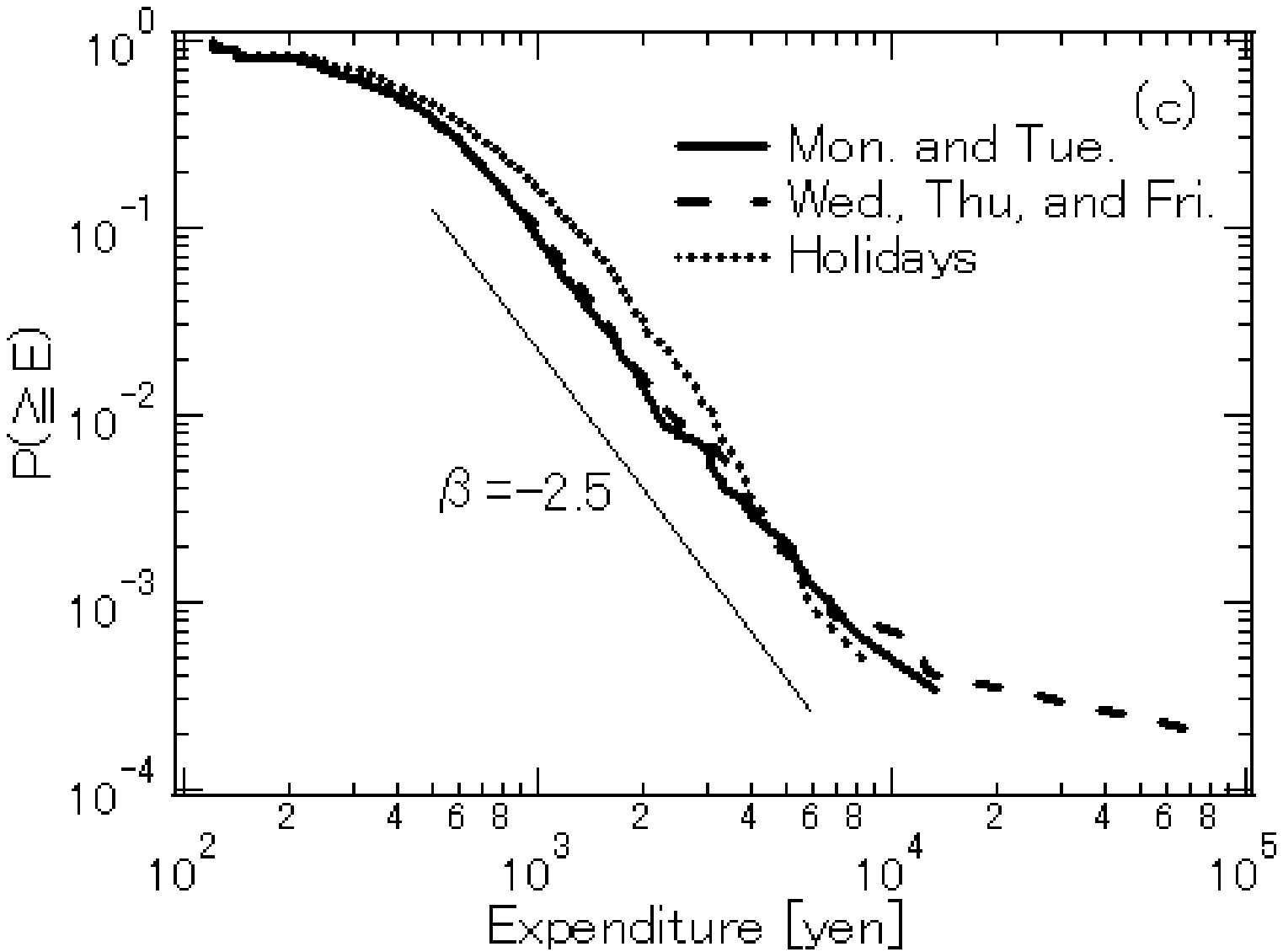}}
\centerline{\includegraphics[height=2.2in]{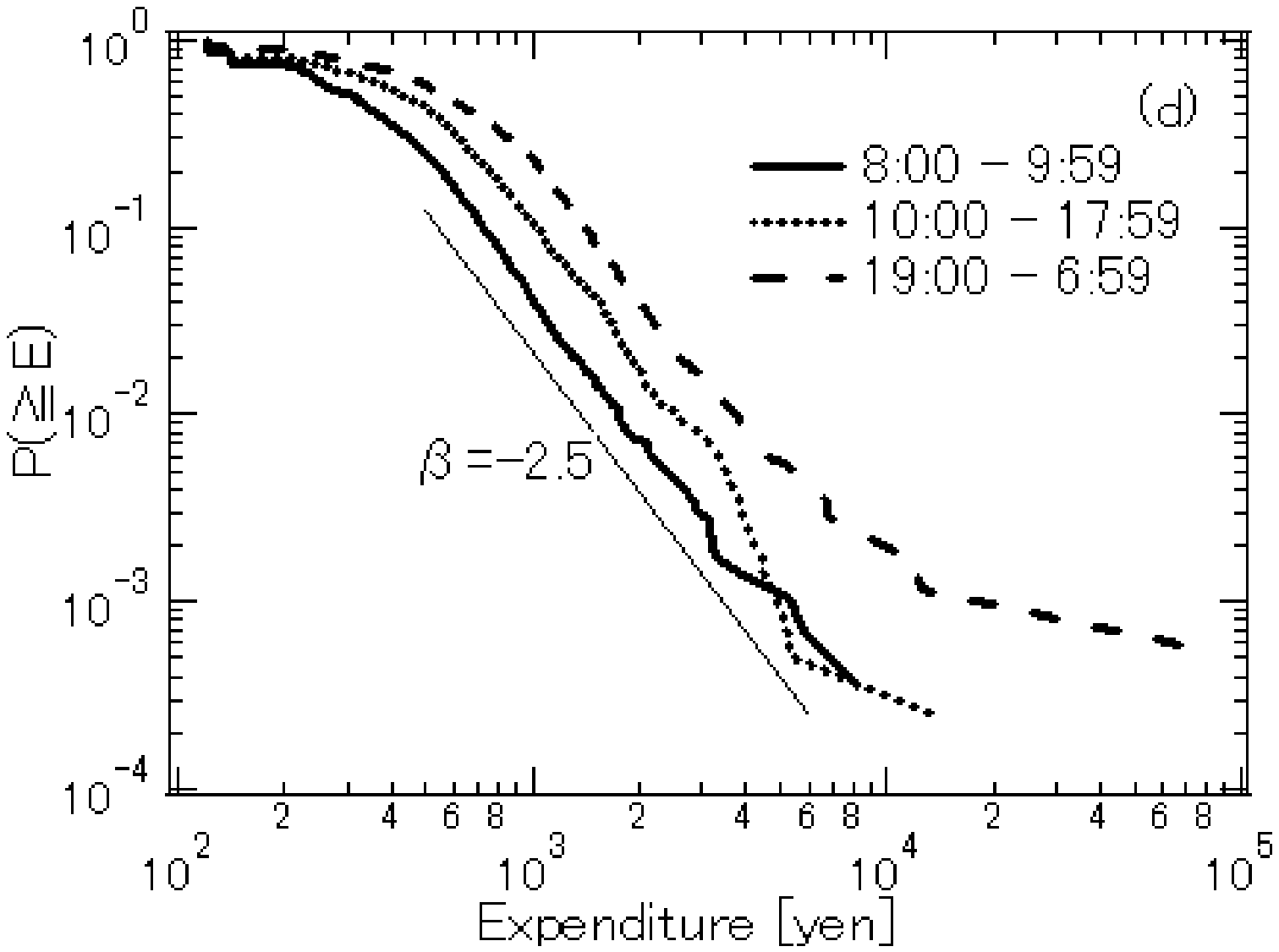}}
\caption{(a), (b), (c), (d)}
\end{figure}

\end{document}